%% file: ieee_comsoc_main.tex
\documentclass[journal,comsoc]{IEEEtran}
\usepackage[printonlyused]{acronym}
\usepackage[T1]{fontenc}
\usepackage[style=ieee,backend=bibtex]{biblatex}
\bibliography{biblio2}
\usepackage{amsmath}
\usepackage[cmintegrals]{newtxmath}
\usepackage{hyperref}
\usepackage{graphicx}
\usepackage[font=small,justification=centering]{caption}
\renewcommand{\figurename}{Figure}
\usepackage[font=small]{caption}
\usepackage{upgreek}

\begin{document}

\title{Toward Location-aware In-body Terahertz Nanonetworks with Energy Harvesting}

\author{Filip Lemic, Sergi Abadal, Aleksandar Stevanovic, Eduard Alarc\'{o}n, Jeroen Famaey
\thanks{F. Lemic and J. Famaey are with Internet Technology and Data Science Lab (IDLab-Antwerpen), University of Antwerp - imec, Belgium, e-mail: \{name.surname\}@uantwerpen.be}
\thanks{F. Lemic, S. Abadal, and E. Alarc\'{o}n are with NaNoNetworking Center in Catalunya (N3Cat), Universitat Politècnica de Catalunya, Spain, e-mail: \{surname\}@ad.upc.edu}
\thanks{A. Stevanovic is with  Faculty of Medicine, University of Belgrade, Serbia, e-mail: aleksandar.stevanovic@med.bg.ac.rs}\vspace{-1mm}}

\markboth{PREPRINT}%
{F. Lemic \emph{et al.}: Scalable Localization-enabled In-body Terahertz Nanonetworks}

\maketitle

\begin{abstract}
Nanoscale wireless networks are expected to revolutionize a variety of domains, with significant advances conceivable in in-body healthcare.
These nanonetworks will consist of energy-harvesting nanodevices passively flowing through the bloodstream, taking actions at certain locations, and communicating results to a more powerful \ac{BAN}.
Assuming such a setup and electromagnetic nanocommunication in the \ac{THz} frequencies, we propose a network architecture that can support fine-grained localization of the energy-harvesting in-body nanonodes, as well as their two-way communication with the outside world.  
The main novelties of our proposal lie in the utilization of location-aware and \acf{WuR}-based wireless nanocommunication paradigms, as well as \acp{SDM}, for supporting the envisioned functionalities in THz-operating energy-harvesting in-body nanonetworks. 
We argue that, on a high level, the proposed architecture can handle (and actually benefits from) a large number of nanonodes, while simultaneously dealing with a short range of THz in-body propagation and highly constrained nanonodes.   
\end{abstract}


\IEEEpeerreviewmaketitle

\input{acronym_def}
\input{introduction}

\input{architecture}

\input{localization}

\input{communication}

\input{discussion}
\input{conclusion}

\section*{Acknowledgment}

This work received funding via the Horizon 2020 Marie Curie Actions (grant nr. 893760) and Future Emerging Topics (grant nr. 736876).

\renewcommand{\bibfont}{\footnotesize}
\printbibliography

\end{document}

%% file: acronym_def.tex

\acrodef{SDMs}{Software-Defined Metamaterials}
\acrodef{SDM}{Software-Defined Metamaterial}
\acrodef{THz}{Terahertz}
\acrodef{UWB}{Ultra Wide-Band}
\acrodef{FPGA}{Field Programmable Gate Array}
\acrodef{ToF}{Time of Flight}
\acrodef{AoA}{Angle of Arrival}
\acrodef{RSS}{Received Signal Strength}
\acrodef{AP}{Anchor Point}
\acrodef{3D}{3-Dimensional}
\acrodef{TS-OOK}{Time-Spread ON-OFF Keying}
\acrodef{RF}{Radio Frequency}
\acrodef{2D}{2-Dimensional}
\acrodef{GPS}{Global Positioning System}
\acrodef{IoT}{Internet of Things}
\acrodef{WSN}{Wireless Sensor Network}
\acrodef{LoS}{Line of Sight}
\acrodef{BAN}{Body Area Network}
\acrodef{MAC}{Media Access Control}
\acrodef{ML}{Machine Learning}
\acrodef{WuR}{Wake-up Radio}
\acrodef{SNR}{Signal-to-Noise Ratio}
\acrodef{URLLC}{Ultra-Reliable Low-Latency Communication}
\acrodef{D2D}{Device-to-Device}

%% file: introduction.tex
\section*{Introduction}
\label{introduction}

\IEEEPARstart{R}{ecent} developments in nanotechnology are bringing light to nanometer-size devices that will enable a variety of groundbreaking applications. 
In-body healthcare is among many of the interesting domains where nanotechnology is expected to be beneficial. 
Nanotechnology is envisioned to enable molecular-level detection of viruses and bacteria, high-precision drug delivery, targeted monitoring, and neurosurgery~\cite{agoulmine2012enabling}.
For enabling such applications, nanodevices comprising an in-body nanonetwork will flow through the bloodstream, take actions upon commands at target locations, and communicate results to a more powerful \acf{BAN}~\cite{akyildiz2008nanonetworks,dressler2015connecting}. 
In this context, assume the nanodevices with the sizes comparable to the ones of red blood cells (i.e., around 5$\mu$m in diameter), primarily to avoid clotting due to their introduction into the bloodstream.    
Given the small sizes of these nanonodes, harvesting surrounding energy (e.g., from blood currents) is expected to be their sole powering option~\cite{piro2015design}. 
Due to their constrained energy and tiny form factors, these nanonodes are anticipated to be passively flowing, i.e., without the possibility of mechanical steering toward the targeted location.

To support controlling the nanonodes upon reaching their target locations, there is intuitively a need for knowing their current locations. 
There is also a need for communication between the outside world (i.e., BAN) and the nanonode (e.g., for issuing control commands), as well as between the nanonode and the outside world (e.g., for delivering its readings). 
One of the promising enablers for communication in such a scenario is to utilize electromagnetic signals in the \acf{THz} frequencies. 
This is because the communication in these frequencies allows for tiny transceiver form-factors, the prime requirement for in-body nanonodes.
However, the THz band has its peculiarities, primarily pertaining to high scattering and spreading losses, resulting in a low range of in-body THz propagation (i.e., up to a few centimeters). 
Combined with the limited computational and storage resources, as well as constrained powering of the nanonodes relying only on energy harvesting~\cite{dressler2015connecting}, communication between the BAN and the nanonodes is currently challenging to achieve.
The main challenges include i) mitigating the high attenuation of in-body THz propagation, ii) maintaining a low energy profile and complexity of the nanonodes, iii) supporting unprecedented network scalability, and iv) enabling fine-grained localization of the nanonodes~\cite{lemic2019survey}. 

Current research efforts in this direction are sparse, mostly focusing on addressing only some of the above-stated challenges. 
In terms of in-body localization, the existing approaches focus on coarse-grained localization, resulting in a detection of the body region in which a nanonode is located~\cite{dressler2015connecting}, which is not sufficient for many applications (e.g., in diagnostics). 
In terms of communication, current research mostly focuses on random channel access-based link-layer and flooding-based network-layer protocols, both known to increase the interferences in the system, making the envisioned nanonetworks with a large number of nanonodes infeasible in practice~\cite{lemic2019survey}.
Finally, many approaches assume that if the nanonetworking protocols are carefully designed, the energy-harvesting nanonodes could experience perpetual operation, which, we will argue, is impossible in practice.

\begin{figure*}[!t]
\vspace{-1mm}
\centering
\includegraphics[width=\linewidth]{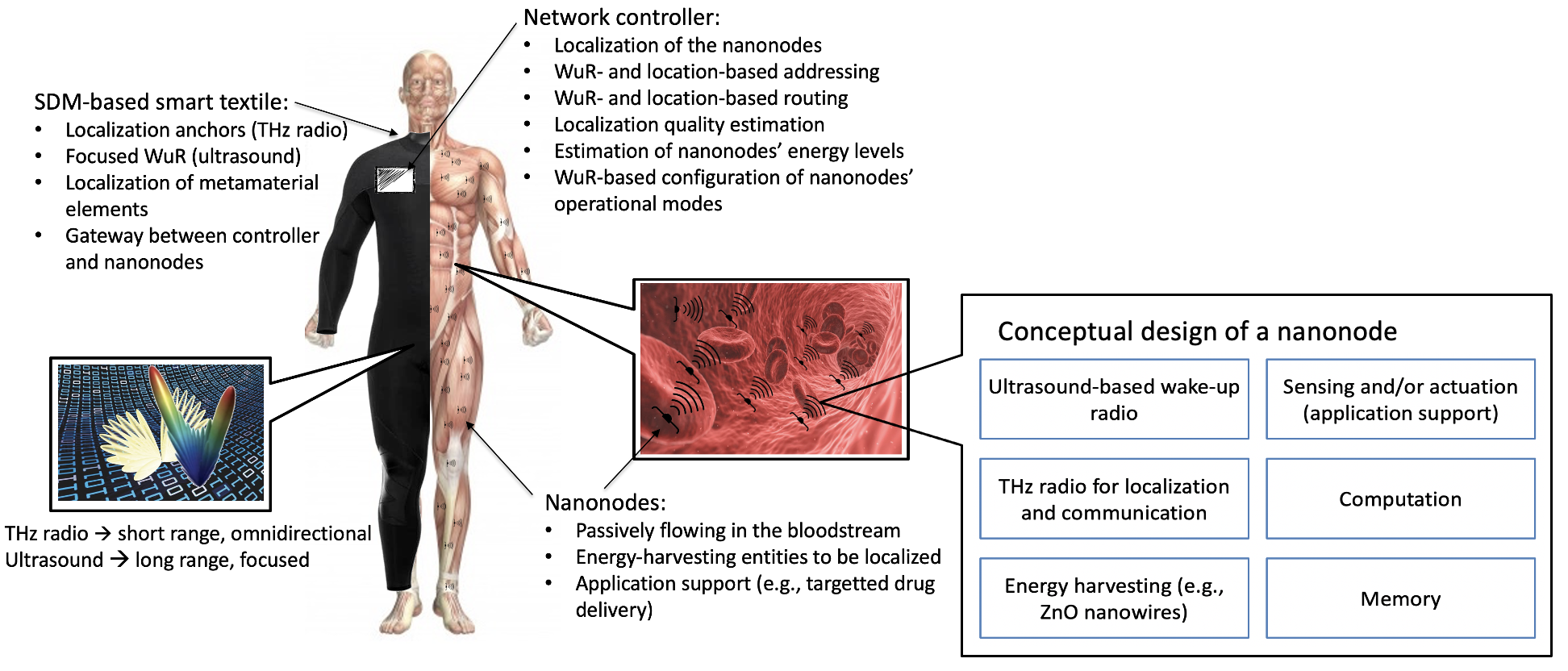}
\vspace{-4mm}
\caption{Conceptual overview of the system}
\label{fig:architecture}
\vspace{-1mm}
\end{figure*}

Based on the shortcomings identified above, we outline a network architecture consisting of a \acf{SDM}-based \ac{BAN}, a \ac{ML}-based network controller, and a large number of energy-harvesting nanonodes flowing in the bloodstream, with the high-level functionalities of the components given in Figure~\ref{fig:architecture}.
Furthermore, we propose an approach for localizing the nanonodes using THz signals without the need for surgically implanted localization anchors, as the localization procedure can be based on utilizing solely the \ac{SDM}-based \ac{BAN} nodes.
Moreover, we propose the usage of a directional nanoscale \acf{WuR} based on ultrasound (hence easily penetrating the human body) for location-based wake-up of only a desired set of nanonodes.
Location-based route-selection is envisioned to support communication between the nanonodes and the \ac{SDM}-based BAN, which in turn benefits the energy consumption of the nanonodes (compared to flooding). 
Finally, the network controller serves to estimate the quality of localization, as well as the energy levels of the nanonodes, which is then used for optimizing the selection of the nanonodes to be awoken, resulting in an optimized selection of the communication paths between the nanonodes and the outside world.

%% file: architecture.tex
\section*{Location-Aware Terahertz-operating In-body Nanonetworks with Energy Harvesting}
\label{architecture}

The authors in~\cite{lemic2019survey} discuss the requirements that in-body healthcare applications are likely to pose on the supporting nanonetworks, which are summarized in Table~\ref{tab:applications_requirements} with the indication of whether or not they are considered in this work. 
With the design requirements in mind, the high-level network architecture, as well as a block diagram with the main components of a nanodevice, are depicted in Figure~\ref{fig:architecture}.
The nanodevice consists of a module for harvesting environmental energy, with the most promising approaches for such harvesting at the nanoscale based on vibrational cycles of ZnO nanowires generated by blood currents or heartbeats, or \ac{RF} power transfer~\cite{piro2015design}. 
The nanodevice also features modules for computation, data storage, and application support, as indicated in the figure.
Application support can be defined as sensing and/or actuation functionalities that the nanodevice is envisioned to enable, for example detection and localization of cancer cells (sensing example) or targeted drug release (actuation example).
These functionalities are well-established in the existing literature~\cite{canovas2018nature}.

\begin{table}[!t]
\begin{center}
\caption{Summary of requirements that healthcare applications are expected to pose on the supporting nanonetworks}
\label{tab:applications_requirements}
\begin{tabular}{p{3cm} p{2.8cm} p{1.2cm}} 
\hline
\textbf{Requirements} & \textbf{In-body healthcare} &  \textbf{Considered?} \\ 
\hline
Network size       & 10$^3$ to 10$^{9}$ & \hfil \checkmark \\
Node density       & $10-$10$^3$ per cm$^3$ & \hfil \checkmark \\
Latency            & ms to s &  \\
Network throughput & 1-50~Mbps & \\
Traffic type       & bidirectional & \hfil  \checkmark \\ 
Reliability  	   & very high & \hfil  \checkmark \\ 
Energy consumption & very low & \hfil  \checkmark \\
Mobility           & high &  \hfil \checkmark \\
Addressing         & individual & \hfil \checkmark \\
Security           & very high & \\
Additional features & localization \& tracking & \hfil \checkmark \\
\hline
\end{tabular}
\end{center}
\vspace{-3.5mm}
\end{table}

In the context of this work, the most interesting modules are the \ac{THz} radio for localization and data communication, and the \ac{WuR} used for the wake-up of the nanonodes.
From the technological point of view, a THz-operating transceiver can be implemented at the nanoscale by utilizing graphene, which for low range communication could theoretically feature extremely high bandwidths (in the order of hundreds of Gigahertz (GHz)) for short distances of roughly 1~cm~\cite{jornet2013graphene}.

On the physical layer we consider the usage of time-spread ON-OFF keying (TS-OOK), as it is a de-facto standard communication scheme for nanocommunication in THz frequencies.
In TS-OOK, very short pulses (i.e., 100~fs long) represent logical “1”s, while logical “0”s are represented by silences~\cite{jornet2014femtosecond}. 
The time between the transmission/reception of two consecutive bits of a packet is characterized by $\beta$ and generally much longer (i.e., two to three orders of magnitude) than the duration of a TS-OOK pulse.
This is a feature of TS-OOK allowing for unsynchronized transmission with low chance of interference.  
The energy consumption modeling in this scheme has usually been carried out by attributing certain consumed energy to transmission and reception of a TS-OOK pulse. 
However, one could argue that, from more traditional wireless networking approaches such as \acp{WSN}, we know the energy consumption of any wireless transceiver is dominated by transmission, reception, and \textit{idling}. 
This argument has been outlined in~\cite{lemic2020idling}, where the authors consider idling energy in the overall energy consumption of a THz-operating TS-OOK-based nanonode, in addition to the energies consumed in transmission and reception. 
They show that such nanonodes will mostly not have enough energy for communication, which is a direct result of the long idling time $\beta$ between the transmission/reception of consecutive pulses or silence, and (more importantly) the long expected time between transmission/reception of consecutive packets for the expected case of low data-rate communication.  

\begin{figure}[!t]
\centering
\includegraphics[width=\linewidth]{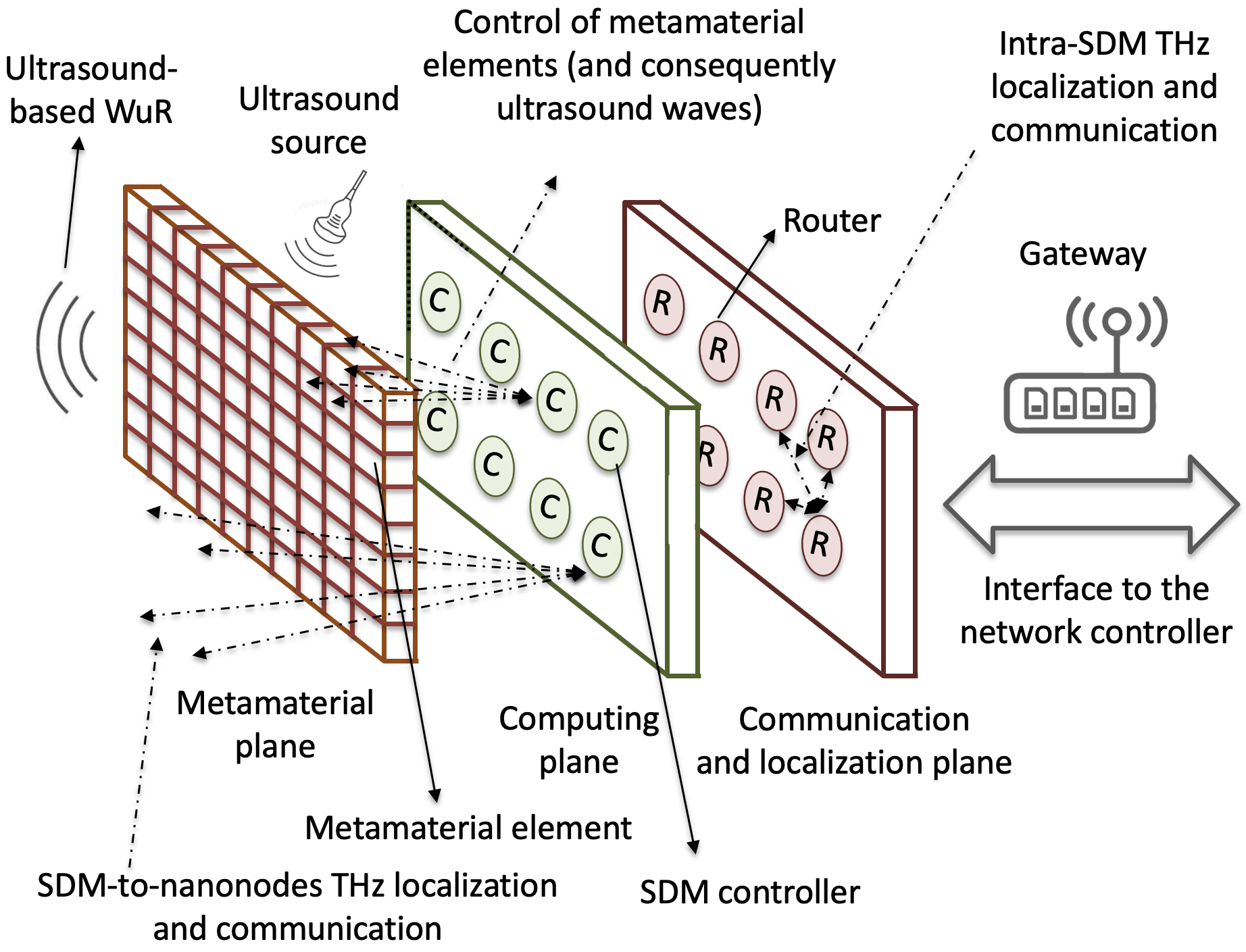}
\caption{\acf{SDM} architecture}
\label{fig:sdm_architecture}
\vspace{-1mm}
\end{figure} 

Based on~\cite{lemic2020idling}, we argue that the energy-harvesting nanonodes will have to be asleep most of the time in order not to deplete their energy levels in idling.  
Therefore, there is intuitively a need for their wake-up when they are needed, which is in our design of a nanodevice (Figure~\ref{fig:architecture}) enabled through the \ac{WuR}.
Specifically, we envision the utilization of an ultrasound-based and, therefore, human body-penetrable nanoscale \ac{WuR}.
The design and implementation of one such system have been presented in~\cite{simeoni2020long}, which is based on a piezoelectric Nanoscale Ultrasound Transducer (pNUT) generating a voltage signal by picking up pressure changes due to incident ultrasound.  
The authors demonstrate the WuR is able to maintain robust data transfer over a range of 0.5~m when operating with a 40~kHz carrier signal modulated at 250~Hz, which intuitively would suffice for reaching the nanonodes in any region inside the human body.  
With its sizes of 30$\upmu$m$\times$30$\upmu$m$\times$150\hspace{0.1mm}nm~\cite{simeoni2020long}, bringing it down to the required size of several micros is conceivable, given that it is envisioned to be used for waking up the nanonodes and configuring their operational mode, in contrast to technologically more challenging full-scale two-way communication with the outside world.  

On the \ac{BAN} level, we envisage the use of an \ac{SDM} covering the human body in the form of a smart textile. 
Metamaterials feature engineered properties that can be exploited for deep-sub-wavelength manipulation of ultrasound waves (e.g., reflection, absorption, scattering), both in near and far-fields~\cite{chen2018deep}.
\ac{SDM} is a paradigm for enabling a substantially higher level of controllability of metamaterial elements in real-time, allowing for their reusability across applications and operations~\cite{liaskos2018new}.
It is worth pointing out that the SDMs are envisioned to be flexible entities with possibilities of bending, stretching, and rolling~\cite{walia2015flexible}. 
Therefore, they will be embeddable in smart textiles and operational even under structural changes, with their thickness for required for controlling ultrasound waves operating at 40~kHz being in the range of a few millimeters~\cite{chen2018deep}.
This implies that the envisioned network architecture would be feasible even in case of an environment with mobility (i.e., the person carrying the system would not have to be immobilized).  
We envision an \ac{SDM} architecture as depicted in Figure~\ref{fig:sdm_architecture}, in which the SDM is used i) for controlling ultrasound waves to serve as focused \ac{WuR} signals for waking up the nanonodes in the body, ii) as anchors for THz-based localization of the nanonodes, and iii) as data relays between the nanonodes and network controller.
As shown in the figure, the SDM consists of metamaterial elements with their control being enabled through THz-based nanocommunication, which is in line with with the existing literature~\cite{liaskos2018new,lemic2020thz_loc}.
The same type of THz nanocommunication is used for communication and localization of the nanonodes in the body.   

ass the THz-based nanocommunication transceivers used for controlling the metamaterial elements are also envisioned to serve as potential anchors for localization of the in-body nanonodes (cf., Figure~\ref{fig:sdm_architecture}), they are expected to either be fixed on the immobilized body or (more realistically) localizable relative to each other in case of the deformations of SDMs due to the body movements. 
In this direction, it is worth stressing that early approaches exist to internally localize the metamaterial elements and corresponding THz transceivers used for their control, implying that the locations of BAN nodes on the body can be derived even under mobility~\cite{lemic2020thz_loc}.

%% file: localization.tex
\section*{Localization in Terahertz-operating Energy-harvesting In-body Nanonetworks}
\label{localization}

Localization can be considered as the first (also mandatory) phase in the operation of the proposed architecture.
The proposed approach in enabling localization is grounded in the work by Slottke~\cite{slottke2016inductively}.
In~\cite{slottke2016inductively}, Slottke envisions a sensor network consisting of hundreds or thousands of wireless micro-nodes with sub-millimeter dimensions.
For such a setup, Slottke shows that the micro-nodes can be accurately localized by utilizing inductive coupling between them and the anchors. 
He demonstrates that such localization can be performed iteratively, so that the "range" of localization is increased. 
Iterative localization implies first localizing the micro-nodes close to the skin (i.e., where there is a sufficient number of localization anchors), followed by utilizing (some of) the newly localized micro-nodes as virtual anchors in the following iterations.
In this case, a relatively large number of, ideally optimally positioned, anchors (either real or virtual) is required for maintaining good localization performance, which might be infeasible in practice, argues Slottke. 

\begin{figure}[!t]
\centering
\includegraphics[width=\linewidth]{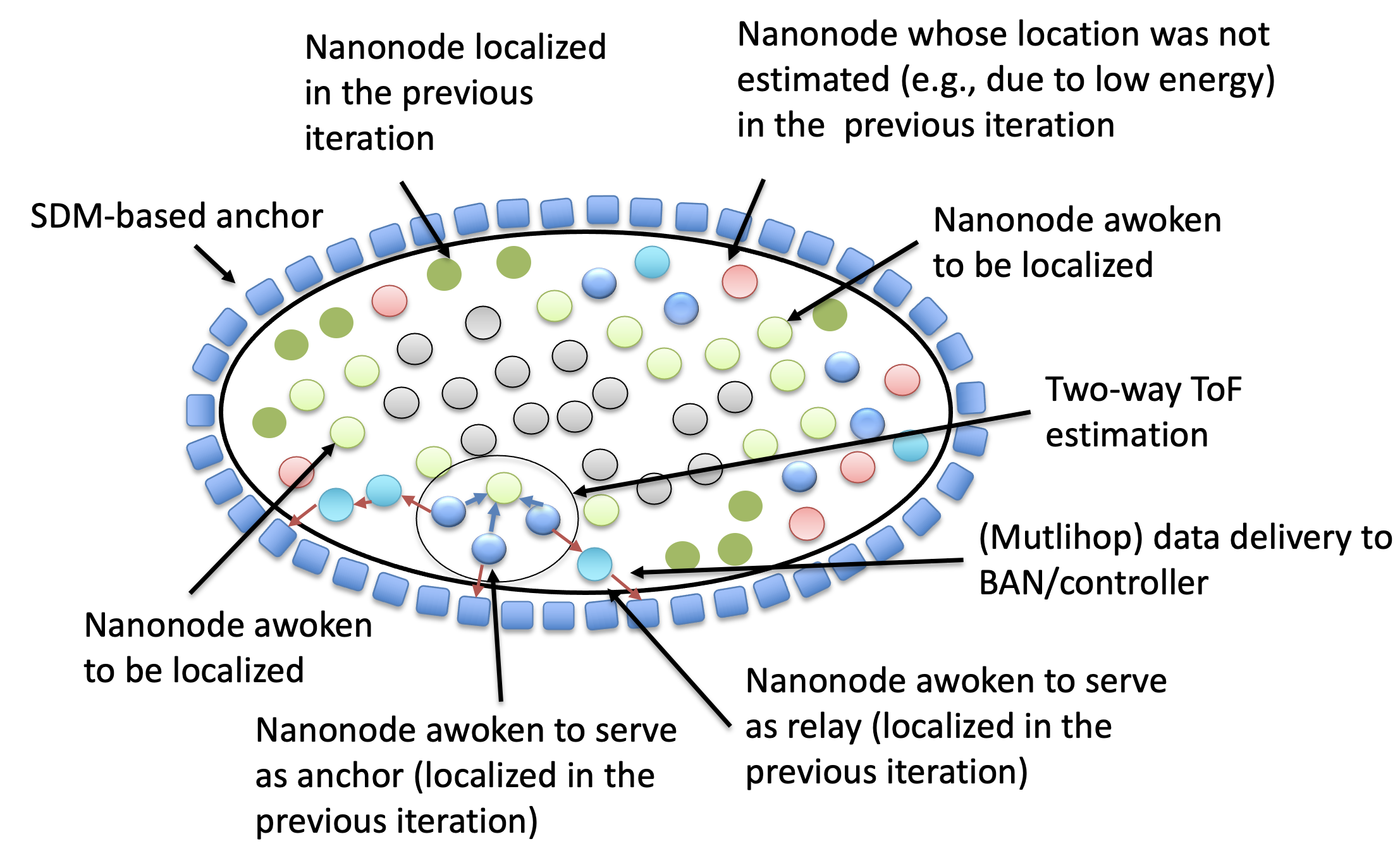}
\caption{Iterative localization scheme}
\label{fig:iterative_localization}
\vspace{-1mm}
\end{figure}

Although the outlined approach is not directly applicable at the nanoscale, it provides several guidelines on how nanoscale localization inside of the human body could be performed. 
First, it is important to realize that in-body nanonetworks will have to be extremely dense for many of the envisioned applications (this could also be considered as a design requirement for our system), suggesting that the accuracy of localization could be very high.
Second, although Slottke states that \ac{ToF} estimation cannot be performed due to the narrowband nature of the wireless communication candidates considered in his work, at THz frequencies for small transmission distances of up to a few centimeters that is not the case.
In contrast, for such distances and THz-operating graphene antennas, even THz-wide nanocommunication can be achieved~\cite{lemic2020thz_loc}.
This implies that highly accurate two-way ToF estimation between metamaterial elements and the nanonodes, as well as among the nanonodes, can be performed.
Assuming 1~THz of available bandwidth, the raw resolution of such sampling would result in a mm-level ranging errors, which would roughly translate to the localization accuracy of the same order of magnitude, assuming localization in a \ac{3D} space and minimum required number of anchors (i.e., four). 
Note that two-way ToF estimation does not require tight synchronization between the devices, which would be complex to achieve for the considered setup. 
Also note that, though 1~THz of bandwidth (implying comparable sampling frequency at the nanonode level) is not practically feasible, different approaches exist for two-way ToF sampling that can mitigate this limitation and effectively result in bandwidth-bound (in contrast to sampling-bound) two-way ToF ranging errors (more details in~\cite{lemic2020thz_loc}).  

Slottke's argument about the need for a large number of (ideally optimally positioned) localization anchors provides the justification for the utilization of a body-worn \ac{SDM} containing such anchors. 
The utilization of graphene-based THz signals in the localization process implies tiny transceiver sizes (e.g., a 1-micron large graphene-based transceiver operating at 1~THz has been conceptually outlined in~\cite{lemic2019survey}), which in turn suggests that many potential localization anchors could be embedded in the smart textile, providing an unprecedented degree of freedom in the optimization of localization in terms of number and locations of the anchors. 
We argue that the above considerations illustrate the potential of THz-, two-way ToF-, and SDM-based localization of in-body nanonodes. 

The envisioned localization setup is given in Figure~\ref{fig:iterative_localization}.
In the initial step (not depicted), the \ac{SDM}-based anchors are to be localized, in case their locations are not fixed and known. 
For localizing the anchors one can utilize the localization approach proposed in~\cite{lemic2020thz_loc}, which is comparable to the one outlined here (i.e., two-way ToF-based trilateration), with the differences being the propagation environment (free-space vs. in-body) and iterative nature of localization. 
In the second step, the anchors are envisioned to perform two-way ToF estimation with the nanonodes, once these nanonodes are awoken using ultrasound-based \ac{WuR}.
The awakening should be done for all the nanonodes close to the skin and, therefore, in direct communication range with multiple anchors.
In the subsequent step, some of nanonodes whose locations are now estimated can become ``virtual'' anchors in the localization of the nanonodes deeper inside the body while others can become data relays toward the outside world, as depicted in the figure. 
Intuitively, the nanonodes should be awoken in order i) to be localized, ii) to serve as virtual anchors in localization, iii) to serve as relays in delivering the data to the outside world (for both localization and communication), and iv) to serve the application purposes (e.g., initiating sensing or actuation and communicating the results or success of such actions). 
This four-states level of control of an individual nanonode can be achieved by transmitting carefully designed sequences of directional \ac{WuR} signals.  


%% file: communication.tex
\section*{Communication in THz-operating Energy-harvesting In-body Nanonetworks}
\label{communication}

As mentioned, the location estimates will feature certain localization errors, which will be higher for the estimates deep inside the body compared to the ones closer to the SDM-based anchors.
In the following, we will assume the locations of the nanonodes in a given time instance are estimated and the qualities (characterized by the expected level of errors in localization) of such estimates are known. 
Along these lines, we postulate that the location-aware wireless communication paradigm is a strong candidate for enabling two-way communication between the outside world and the in-body nanonodes, which will be demonstrated in this section. 

\begin{figure}[!t]
\centering
\includegraphics[width=\linewidth]{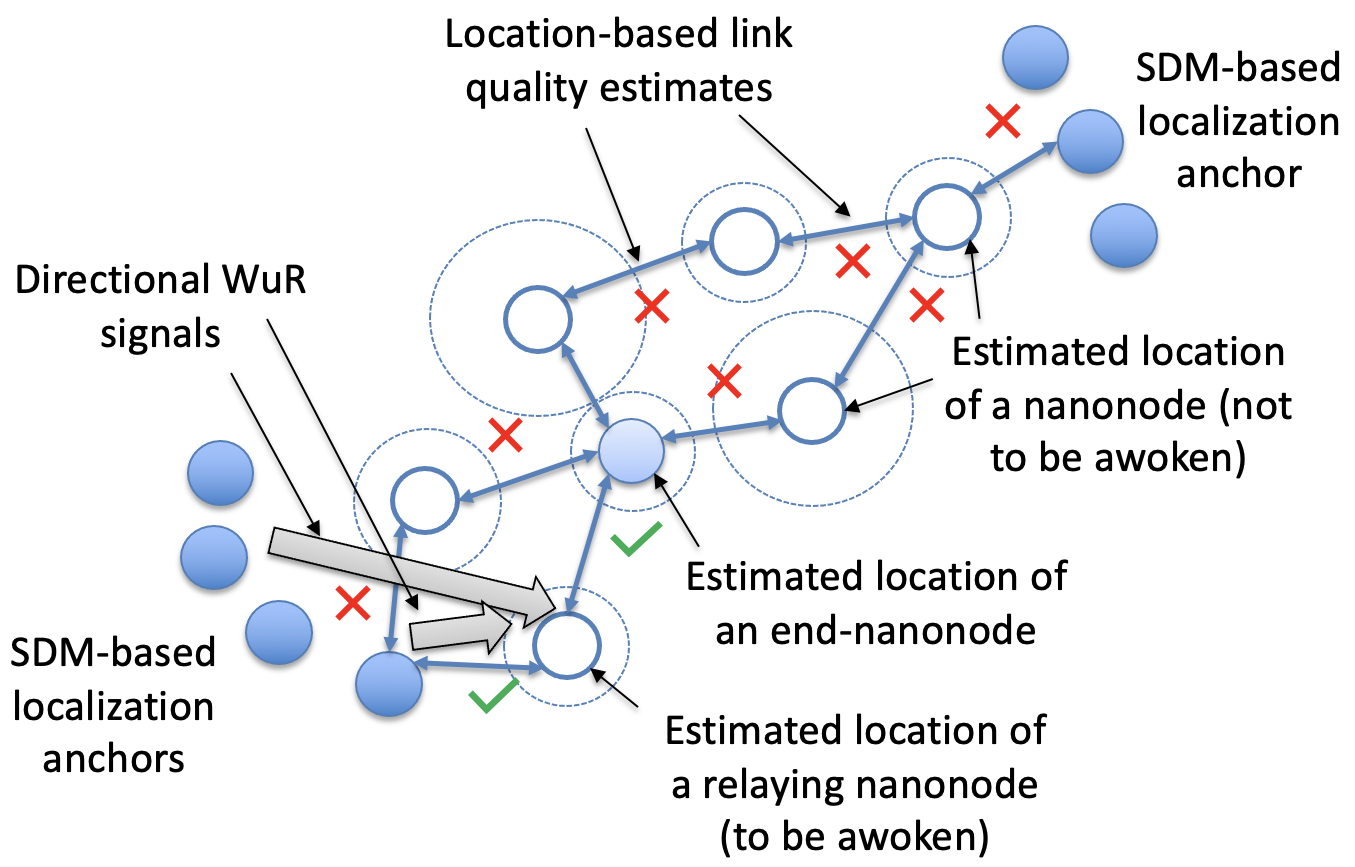}
\caption{Location-based wake-up and route selection}
\label{fig:wake_up}
\vspace{-1mm}
\end{figure}

There are two main communication-related aspects considered in this work, i.e., addressing and routing. 
In terms of addressing, we posit that the nanonodes can be awoken and, therefore, addressed individually, if their locations can be estimated and if the quality of such estimation can be obtained. 
Our high-level approach in waking up the nanonodes is based on issuing directional WuR signals from the SDM, as indicated in Figure~\ref{fig:wake_up}.
Specifically, by utilizing an SDM one could sequentially transmit multiple directed WuR signals in such a way that only the nanonode(s) to be awoken receive(s) all of them.
By doing that, one could wake up only a desired nanonode and, therefore, achieve individual addressing of that nanonode based on its practically-relevant location. 
Moreover, by waking up only desired nanonodes, in contrast to waking up all of them in a given region, one could create a multi-hop routing path for data delivery from a particular nanonode to the outside world, for both localization and communication. 

Existing literature provides a variety of approaches for location-based link quality estimation for the purposes of \ac{D2D} link establishment, positioning of a mobile relay, handover between base stations or network technologies, etc. 
Though these approaches are usually outlined for macro-scale wireless networks, some of them are interesting in this context as they assume the availability of \textit{erroneous} location information, where the levels of average localization errors can be quantified.  
Based on such approaches, we can outline, on a high-level, our approach for location-based (nb., and location quality-based) route selection.
Specifically, a metric characterizing the expected link quality (e.g., expected \ac{SNR}) between two nanonodes can be established based on their estimated locations, expected errors of these estimates, their estimated energy levels, and high-level knowledge about the path-loss in the environment (cf., Figure~\ref{fig:wake_up}).
The metric can then be used for selecting the nanonodes to be awoken based on the WuR approach outlined before. 
Note that the mentioned approaches can be used in case of both devices featuring a localization error, as well as when the location of one device is perfectly accurate, which is intuitively suitable for all cases envisioned in this work (i.e., anchors' locations are known or can be estimated with a certain level of accuracy, while nanonodes' location estimates always feature a localization error).

%% file: discussion.tex
\section*{Performance Results and Discussion}
\label{discussion}

\begin{figure}[!t]
\centering
\includegraphics[width=\linewidth]{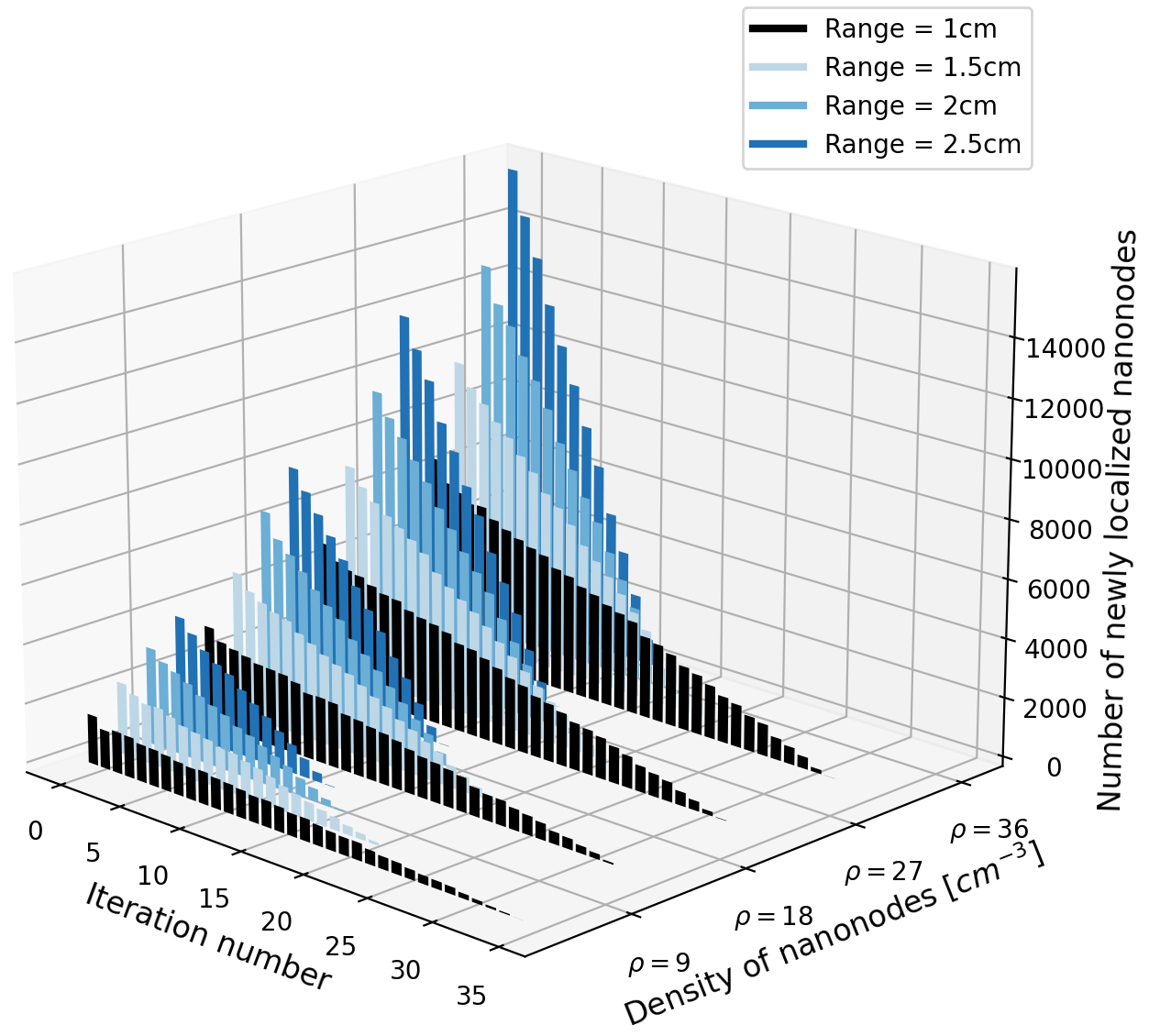}
\caption{Number of localization iterations as a function of the communication range and nanonode density}
\label{fig:result}
\vspace{-1mm}
\end{figure}

In Figure~\ref{fig:result}, we depict the number of nanonodes localized in each iteration as a function of their communication range and density in the human body. 
We do that for an environment representing a 1~cm thick slice of the torso area, approximated as a circle with the radius of 30~cm. 
We assume there is a substantially large number of SMD-based anchors for localization, primarily due to their achievable small sizes (e.g., 1 micron as reported in~\cite{lemic2019survey}).
Hence, we assume that the randomly distributed nanonodes can be localized in the first iteration if their distances to the edge of the circle are smaller than their communication range.
The accuracy that can be achieved in the first iteration is modeled as a zero-mean Gaussian variable $\mathcal{N}(0,\sigma^2)$.
In the following iterations, due to the fact the considered environment is a slice of a realistically expected one, the nanonodes are assumed to be localizable if their distances to three already localized nanonodes are smaller than the achievable communication range. 
In the $n^{th}$ iteration, the compounding localization errors can be approximated by $n\sigma^2$. 
Such an approximative approach for simulating iterative localization for an extremely large number of entities has been utilized in the existing literature (e.g., BitSimulator).
Other simulation parameters are consistent with~\cite{lemic2020thz_loc}.

As visible from the figure, for the expected cm-level communication ranges and nanonode densities aligned with the smallest expected nanonode densities from Table~\ref{tab:applications_requirements}, the localization can generally be performed in less than 35 iterations.
Given the compounding nature of localization errors in each iteration and assuming mm-level average localization accuracy based on~\cite{lemic2020thz_loc}, the average localization errors of the proposed scheme in the considered experiment are statistically bound by 3.5~cm.
In addition, Figure~\ref{fig:result} shows that this statistical error bound can be further reduced by increasing the density of nanonodes, demonstrating that the proposed scheme actually benefits from this increase.

We assert that the outlined localization approach intuitively addresses the issues stemming from a short range of THz in-body propagation under the assumption of high density of anchors and nanonodes. 
We view the approach as suitable for energy-harvesting nanonodes, primarily because it is based on the exchange of only a small number of TS-OOK pulses and does not require tight synchronization among the nanonodes nor between the nanonodes and the anchors, and can be supported by simple transceiver architectures~\cite{jornet2013graphene}. 
Moreover, we argue that our approach addresses the infeasibility of flooding-based routing for energy-harvesting nanonetworks, and enables location-based addressing of the nanonodes.
Nevertheless, there are open challenges to be resolved to fully demonstrate the feasibility of the proposed approach.  


\textbf{Multiple Responses:} As numerous nanonodes are expected, their near-simultaneously retransmitted TS-OOK pulses will have to be distinguished at the anchor or controller level, so that their locations can be accurately estimated. 
We hypothesize that, in this regard, the best way forward is to consider random back-offs, as well as trilateration constraints.
In other words, the fact the TS-OOK uses a pulse interval $\beta$ much larger than the pulse duration could be utilized for embedding the retransmissions from multiple nanonodes through random back-offs.
Hence, the chances of multiple responses can be reduced, though the question on the maximum number of retransmissions that could be distinguished in the duration of one interval $\beta$ remains open. 
Note that this number has to be limited by the ability to time discriminate arrival of retransmitted TS-OOK pulses, so that the anchors are able to distinguish two different responses based on the delay between them, accounting for minimum differences in the back-off times and times needed for the signals to propagate through the medium.
In case of multiple signals that cannot be distinguished using solely random back-off, one could aim at the estimation of nanonodes' locations based on all combinations of the signals received by the anchors, and discarding the ones that are estimated to be outside of the body or in the regions covered in previous iterations (i.e., trilateration constraints).      

\textbf{Performance of Iterative Localization:} 
The errors in the scheme compound as a function of the increase in the number of iterations and, therefore, also as a function of the distance between the BAN and the nanonodes (cf., Figure~\ref{fig:result}), resulting in the localization performance worsening in the deeper body parts.  
In the end, the acceptability of a certain localization accuracy will primarily be judged in relation to the requirements of the application(s) to be supported.  
If the proposed approach eventually cannot meet the accuracy posited by applications, one could resort to introducing additional anchors at strategic places inside the body, though this is ideally to be avoided.
Nonetheless, it is an open question on where to position such anchors, which should be approached simultaneously from the localization (i.e, to optimize the localization performance) and surgical perspectives (i.e., to minimize the complexity of the surgical procedure). 
It is also interesting to observe that, in the proposed localization system, multiple subsystems can be specified based on the body regions where the SDMs are mounted.
Hence, potentially multiple iterative localization procedures could be initiated simultaneously (e.g., a number of regions can be specified in the torso area, which requires the highest number of iterations and is, therefore, the most challenging to localize).
In that sense, one could simultaneously obtain multiple estimates (i.e., one from each localization subsystem) of a single nanonode, which could then be used for enhancing the localization accuracy for the nanonodes positioned deep in the body. 
The same approach could be used for scalability or reliability-related enhancements.




\textbf{Reliability:} 
Figure~\ref{fig:result} shows that the number of nanonodes localized in each iteration reduces with the increase in the number of iterations.
This implies that there are always enough nanonodes to be utilized as virtual localization anchors or data relays, compared to the number of nanonodes to be localized in a given iteration, suggesting the reliability of the proposed approach could be high. 
Nonetheless, the question of the level of reliability the proposed approach can achieve remains open. 
In this context, the reliability has to be considered jointly with the latency of data delivery. 
This is primarily because the nanonodes are envisioned to be passively flowing in the bloodstream, thus it is rather important to deliver desired control commands to initiate certain actions at an appropriate time, so that the actions can be executed at a target location.
However, this interplay between nanonodes' (and potentially anchors') mobility on the one hand, and the reliability and latency of data delivery on the other is yet to be investigated.

\textbf{Network Control:} The proposed communication system relies on the estimated locations of the nanonodes, and should intuitively account for the localization errors, especially given the iterative localization process.
This implies that there is a need for estimating the quality of location estimates, ideally on a per-estimate basis. 
We are aware of only one such attempt (discussed in~\cite{lemic2020thz_loc}), where the authors propose an ML-based system for estimating localization errors on a per-estimate basis using solely the raw data (i.e., \ac{RSS}) utilized for generating location estimates. 
Nevertheless, the proposal in~\cite{lemic2020thz_loc} is focused on the macroscale and considers a different localization approach, as well as traditional networking technologies and frequencies.
Hence, its utility for the problem at hand is yet to be established. 

\textbf{Nanonode's Energy Level Estimation:} Finally, accounting for the energy levels of the nanonodes in their awakening and route selection is of prime importance, given that the alternative could result in an unnecessary wake-up of nanonodes whose energy levels are nearly depleted. 
To the best of our knowledge, proposals for such modeling do not exist for the nanonodes assumed here (cf., Figure~\ref{fig:architecture}).
We envision this functionality to be supported by the network controller, together with the discussed functionality of estimating the quality of localization. 
Moreover, we argue that the energy level modeling should be based on the region in which a nanonode is located at a given moment.
Based on the nanonode's location one could postulate the recently harvested amount of surrounding energy.
For example, the nanonode's amounts of harvested energy would intuitively increase as the nanonode moves closer to the heart, assuming that the energy is harvested from the heartbeats.  
One could also aim at utilizing localized historical information for nanonode's energy level characterization. 
Specifically, if one would be able to characterize that in a relatively recent time-frame there was very little activity for the nanonodes in a given region, one could argue the energy levels of the nanonodes in the region are high, hence these nanonodes could be awoken more successfully.
These considerations again demonstrate several benefits of location-aware nanonetworks in this context. 

%% file: conclusion.tex
\section*{Conclusion}
\label{conclusion}

We have outlined an architecture for in-body THz-operating energy-harvesting nanonetworks.
This has been done with the goals of enabling localization of the passively flowing and energy-harvesting nanonodes and their two-way communication with the outside world.
The architecture can handle a large number of nanonodes and very short range of THz in-body propagation.  
We consider it to also be suitable for medical nanorobots in enabling their navigation and communication.   

The envisioned applications pose extreme requirements on the supporting nanonetworks, among others in terms of scalability, energy consumption, reliability, and features such as localization and tracking. 
To meet these unprecedented requirements, we see location and generally context-awareness as a promising network optimization tool. 